# Recommendation systems in e-commerce applications with machine learning methods

Aneta Poniszewska-Marańda*
Magdalena Pakuła*
Bożena Borowska*
aneta.poniszewska-maranda@p.lodz.pl,234766@edu.p.lodz.pl,bozena.borowska@p.lodz.pl
Institute of Information Technology, Lodz University of Technology
Łódź, Poland

## Abstract

E-commerce platforms are increasingly reliant on recommendation systems to enhance user experience, retain customers, and, in most cases, drive sales. The integration of machine learning methods into these systems has significantly improved their efficiency, personalization, and scalability. This paper aims to highlight the current trends in e-commerce recommendation systems, identify challenges, and evaluate the effectiveness of various machine learning methods used, including collaborative filtering, content-based filtering, and hybrid models. A systematic literature review (SLR) was conducted, analyzing 38 publications from 2013 to 2025. The methods used were evaluated and compared to determine their performance and effectiveness in addressing e-commerce challenges.

## 1 Introduction

Recommender systems, also known as recommendation systems (RSs), are a type of information filtering system that employs software tools and techniques to offer suggestions tailored to users' needs. As the volume of information on the internet continues to grow, recommendation systems have become valuable tools across various fields, effectively addressing the challenge of information overload. Recommendation systems are not only utilized in e-commerce to personalize offers but also in other aspects of life such as health industry, education, entertainment, social media, and news platforms. By analyzing patterns in data, recommendation systems improve user experiences by offering highly targeted suggestions and enhancing customer satisfaction. Recommender systems (RSs) have become indispensable in modern digital ecosystems, addressing the growing challenge of information overload by providing suggestions aligned with users' interests.

This paper presents the literature systematic review that explores critical research questions central to the development and deployment of RSs, including the effectiveness of machine learning (ML) methods, the challenges of implementing these systems in dynamic environments like e-commerce, and emerging trends shaping their evolution. These research questions aim to provide actionable insights for improving the design and efficiency of RSs, particularly in e-commerce, where personalization and user engagement are crucial for success.

## 2 Background of Recommender Systems

Recommender systems are integral to modern digital platforms, aiming to alleviate information overload by suggesting relevant content or products to users based on their preferences, behaviors, and historical data. Since their introduction in the early 1990s [1], RSs have undergone significant transformations, driven by advances in machine learning and data mining. These advancements have enabled RSs to handle large datasets, provide more accurate recommendations, and adapt to user behavior in real-time.

Foundational techniques like collaborative filtering and content-based filtering initially tackled the challenge of narrowing user choices. However, with the rapid growth of data and computational capabilities, more sophisticated methods have emerged, including hybrid approaches, deep learning models, and reinforcement learning. These advancements have significantly enhanced the scalability, accuracy, and personalization of recommendation systems (RSs), cementing their role as essential tools across various industries.

In the e-commerce sector, RSs have revolutionized how businesses engage with customers, offering personalized shopping experiences that enhance user satisfaction and loyalty. By analyzing user behaviors, purchase histories, and preferences, RSs recommend products that align with individual tastes, thereby making the shopping process more intuitive and enjoyable. This personalization may not only increases user engagement but also drives sales through upselling, cross-selling, and repeat purchases.

---

*The authors contributed equally to this research.

The significance of RSs in e-commerce cannot be overstated, as they play a crucial role in optimizing business outcomes. However, the implementation of RSs in e-commerce comes with its own set of challenges, including managing large-scale data, ensuring user privacy, and addressing biases in recommendations. Emerging trends, such as integrating RSs with conversational agents and leveraging explainable AI, promise to further enhance their capabilities [27].

## 3 Research Methodology

To explore the current solutions in recommender systems and identify prevailing trends, challenges, and threats the following research questions were defined:

**RQ1** – *What are the most effective machine learning methods used in e-commerce recommendation systems and how are they compared in terms of performance metrics such as accuracy, precision, and recall?* – This question aims to identify and evaluate the various machine learning methods currently applied in e-commerce recommendation systems. Understanding the effectiveness of different machine learning techniques, such as collaborative filtering, content-based filtering, deep learning, and hybrid methods, in terms of key performance metrics (accuracy, precision, and recall) is crucial for choosing and optimizing recommendation systems to enhance user experience and drive sales.

**RQ2** – *What are the primary challenges in implementing recommendation systems in e-commerce environments?* – This question seeks to explore the major challenges found during the implementation of recommendation systems in e-commerce settings. Identifying these challenges, is essential for developing strategies to overcome them and improve the deployment and performance of recommendation systems in e-commerce.

**RQ3** – *How do hybrid recommendation systems compare to traditional approaches (e.g., collaborative and content-based filtering) in terms of handling diverse user needs, complex datasets and performance?* – This question focuses on comparing hybrid recommendation systems with traditional methods to evaluate their effectiveness in addressing diverse user requirements and complex datasets. By assessing the advantages and limitations of hybrid systems compared to traditional approaches, this research provides insights into which methods are better suited for specific scenarios in e-commerce, leading to more tailored and effective recommendation strategies.

**RQ4** – *What advancements and trends are shaping the future of recommendation systems in e-commerce?* – This question aims to investigate the latest developments and emerging trends in the field of recommendation systems. Understanding these advancements is vital for staying up-to-date with technological progress and ensuring that e-commerce recommendation systems continue to evolve and improve, thereby enhancing personalization, user satisfaction, and overall system performance.

To refine the results of the database search, specific inclusion (IC) and exclusion criteria (EC) were established (Table 1). To ensure a robust and high-quality review, this study relied on scholarly databases known for their academic credibility, comprehensiveness, and focus on technological research: IEEE Xplore, ScienceDirect, ACM Digital Library. The following research request foe these database was defined: *("recommendation systems" OR "recommender*

**Table 1: Inclusion and exclusion criteria of research in SLR**

| ID  | Inclusion Criteria |
|-----|--------------------|
| IC1 | Studies must address recommender systems in the context of e-commerce or similar industries. |
| IC2 | Research must employ ML algorithms such as collaborative filtering, content-based filtering, hybrid models, or advanced techniques as deep learning, reinforcement learning. |
| IC3 | Papers must report measurable outcomes such as accuracy, precision, recall, or similar metrics to assess effectiveness. |
| IC4 | Studies published in the last ten years. |
| ID  | Exclusion Criteria |
| EC1 | Papers lacking clear methodological descriptions or measurable outcomes. |
| EC2 | Studies not in the English language. |
| EC3 | Preprints from arXiv or similar repositories were excluded if they showed limited citations or relevance to the RQs. |
| EC4 | Duplicate or incomplete studies, such as abstracts without full findings. |

*systems") AND ("e-commerce" OR "online shopping") AND ("machine learning" OR "deep learning" OR "reinforcement learning" OR "hybrid methods" OR "collaborative filtering").*

The selection process for this study involved a systematic and iterative filtering approach to ensure the inclusion of high-quality and relevant research articles. The procedure consisted of 4 distinct stages: initial search, duplicate removal, selection based on title and abstract, and final selection based on full-text review. Querying the databases with specified search string yielded a total of 7,482 papers, distributed as follows: 368 from IEEE Digital Library, 2,500 from ScienceDirect, and 4,598 from ACM Digital Library. Those values directly translate to percentage values of 5% for IEEE Digital Library, 33% for Science Direct and 62% for ACM Digital Library.

Given the vast number of papers, it was necessary to narrow the focus progressively, concentrating on increasingly specific studies based on criteria such as specific techniques and year of publication. Therefore, all duplicates were identified and removed, ensuring that each study in the dataset was unique. The inclusion and exclusion criteria were then meticulously applied to the titles and abstracts of the remaining studies. Studies that did not meet the relevance, methodological, and quality standards were excluded at this stage. Following this, a thorough review of the full texts of the studies that passed the title and abstract screening was conducted. The same rigorous criteria were applied to ensure that only the most relevant and high-quality studies were retained. This filtering process refined the dataset, reducing the initial set of 7,482 papers to 258 in the pre-final stage. Out of these, only 38 studies were selected for inclusion in the systematic review, as they best addressed the research questions and met all the established criteria.

## 4 Results of literature review

For the purpose of this investigation, only studies from 2013 to 2025 were accepted in order to observe the latest research and ideas in implementing recommendation systems (Fig. 1). Out of 38 studies in total, only 3 (8%) were book chapters, while 12 (32%) were conference papers and 23 (60%) were journal articles (Fig. 2).

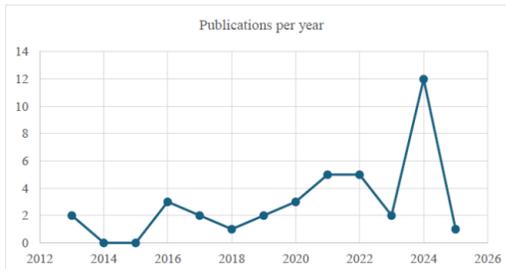

Figure 1: Number of papers selected for SLR by to year.

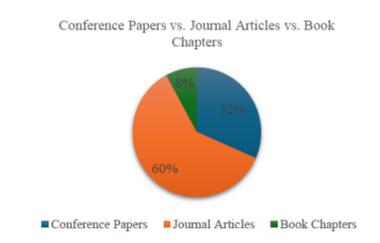

Figure 2: Distribution of studies by publication type.

Table 2: Classification of selected studies according to type of publication

| Classification | Studies | Total |
| --- | --- | --- |
| Journal Articles | [19],[2], [8], [38], [34], [10], [6] [23], [23], [35], [36], [20], [7], [17], [9], [14], [22], [24], [25], [28], [29], [37] | 23 |
| Conference Papers | [39], [15], [4], [3], [12], [26], [30], [18], [11], [31], [16], [32] | 12 |
| Book Chapters | [18], [35], [21] | 3 |

The papers about recommender systems selected for analysis are as follows: [2], [3], [4], [5], [6], [7], [8], [9], [10], [11], [12], [13], [14], [15], [16], [17], [18], [19], [20], [21], [22], [23], [24], [25], [26], [27], [28], [29], [30], [31], [32], [33], [34], [35], [36], [37], [38], [39].

### 4.1 RQ1

The reviewed articles highlight diverse machine learning (ML) techniques for recommendation systems in e-commerce. Popular methods include collaborative filtering, content-based filtering, hybrid approaches, reinforcement learning, and deep learning. These techniques are compared based on their effectiveness using performance metrics such as accuracy, precision, recall, F1-score, MAE, RMSE. Figure 3 presents different types of recommendation system implementations identified, explicitly described and used across the studies.

Based on the studies reviewed, collaborative filtering (CF), content-based filtering (CBF), and hybrid methods came out as the most commonly used techniques. These methods have remained popular throughout the years due to their effectiveness and ability to address different aspects of recommendation tasks.

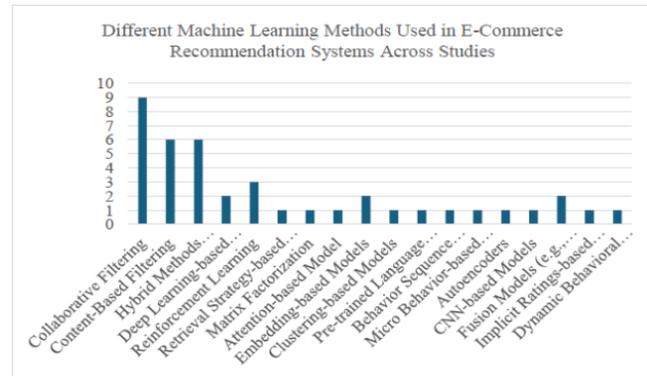

Figure 3: Overview of ML techniques in e-commerce RSs.

Collaborative Filtering is the most frequently employed method, appearing in 9 of the reviewed studies (e.g. [3], [6], [25], [36]). CF is highly favored because it utilizes the preferences of similar users, making it well-suited for e-commerce environments where users often share similar tastes or preferences. Collaborative Filtering suggests products that other similar users have liked, which is particularly effective in large-scale recommendation systems, such as those used by Amazon or Netflix.

Content-Based Filtering and Hybrid Methods are the second and third most used techniques. CBF recommends products based on the features of items that a user has previously interacted with, excelling at recommending items similar to those the user has shown interest in. Hybrid methods, on the other hand, combine both Collaborative Filtering and Content-Based Filtering, aiming to provide more accurate recommendations by mitigating the weaknesses of each individual method.

In addition to traditional methods, Deep Learning-based Models, Reinforcement Learning, and Embedding-based Models are emerging as promising techniques. For instance, Deep Learning (e.g. [8]) shows potential for handling complex data, while Reinforcement Learning (e.g. [19]) adapts recommendations based on user feedback over time. However, these methods are still less commonly applied due to their complexity and high computational requirements.

The increasing use of Hybrid Methods suggests that combining multiple approaches leads to better performance by addressing the weaknesses of individual methods. As computational power increases, methods like Reinforcement Learning and Deep Learning are expected to become more prevalent, offering more personalized and adaptive recommendations. While many studies also present performance metrics such as accuracy, precision, recall, F1-score, MAE, and RMSE, comparing these metrics across different studies can be challenging. Each study uses distinct datasets, which makes direct comparisons difficult, as the quality and characteristics of the data can vary significantly. Moreover, the methods themselves may have been optimized for specific types of data, leading to different performance outcomes even when similar metrics are used.

Therefore, comparing results across studies isn't always ideal. Differences in data, such as variations in size, quality, and domain, can substantially affect the performance of the algorithms. For instance, a method that performs well on a specific dataset may not yield the same results when applied to a different dataset. Despite these differences, it remains a valuable exercise to compare the results, as it provides insights into how well the methods work across different contexts and the relative strengths of various machine learning techniques.

## 4.2 RQ2

Implementing recommendation systems in e-commerce environments includes several key challenges that must be effectively addressed to ensure system effectiveness. One prominent issue is the cold start problem, which arises when new users or items lack historical data, making it difficult to generate recommendations. Hybrid approaches that combine auxiliary data sources, such as demographics and product categories, have been proposed as solutions [21], [3].

Data sparsity is another significant challenge, as insufficient user-item interactions can lead to poor recommendations. To avoid this, hybrid models combining explicit and implicit feedback have been shown to improve performance [22], [2]. Additionally, balancing accuracy with diversity is crucial. Hybrid collaborative filtering techniques help ensure recommendations are both accurate and diverse [24]. Issues like the presence of unhelpful reviews can degrade recommendation quality. Using advanced techniques, such as CNN-based models, helps filter out irrelevant reviews [28]. Overfitting, where algorithms become too tailored to specific datasets, is another common problem, which can be addressed by combining content-based and collaborative filtering methods [29].

Furthermore, continuous improvement is essential for keeping algorithms up-to-date with evolving user preferences, and hybrid systems that allow for this ongoing adaptation have been found effective [34]. Similarly, managing diverse data sources can be complex, but normalization techniques within hybrid systems help integrate various data types [35].

Other challenges include real-time adaptation to dynamic user behavior [19], scalability for large datasets [4], and lack of personalization in traditional collaborative filtering models [23]. Additionally, privacy concerns can be mitigated through hybrid systems that prioritize anonymized or aggregated data [17].

In conclusion, while different studies propose varied solutions to these challenges, their approaches often depend on the specific context, such as the dataset or the type of e-commerce environment. These differences make cross-study comparisons difficult but necessary to understand the full scope of issues and potential solutions (Table 3). Primary challenges in e-commerce recommendation systems and their corresponding difficulty ratings for handling provide a visual overview of varying significance of each challenge (Fig. 4).

## 4.3 RQ3

Hybrid recommendation systems have emerged as a powerful solution to address the limitations of traditional approaches, such as collaborative and content-based filtering. They combine the strengths of multiple algorithms, enabling them to fulfill diverse user needs and complex datasets more effectively. The comparison of various hybrid approaches is summarized in table 4, which highlights their key strengths and weaknesses.

Table 3: Challenges and solutions in e-commerce RSs

| Study | Challenge | Proposed solution |
|---|---|---|
| [21] | Cold Start Problem | Hybrid approach (demographics, content-based data) |
| [3], [22] | Data Sparsity | Hybrid models (implicit & explicit feedback) |
| [24] | Accuracy vs. Diversity | Hybrid collaborative filtering techniques |
| [28] | Unhelpful Reviews | CNN-based models to filter unhelpful reviews |
| [29] | Algorithm Overfitting | Content and collaborative filtering methods |
| [34] | Continuous Improvement | Continuous innovation in hybrid algorithms |
| [35] | Handling Diverse Data Sources | Normalization techniques in hybrid systems |
| [19] | Real-Time Adaptation | Real-time hybrid models |
| [4] | Scalability | Hybrid systems with model-based CF and deep learning |
| [23] | Lack of Personalization | Collaborative filtering with personalized content |
| [17] | Privacy Concerns | Hybrid models with anonymized or aggregated data |
| [6] | Dynamic Data | Hybrid approaches with reinforcement learning |
| [3] | Complex Datasets | Hybrid models combining user behavior and content features |

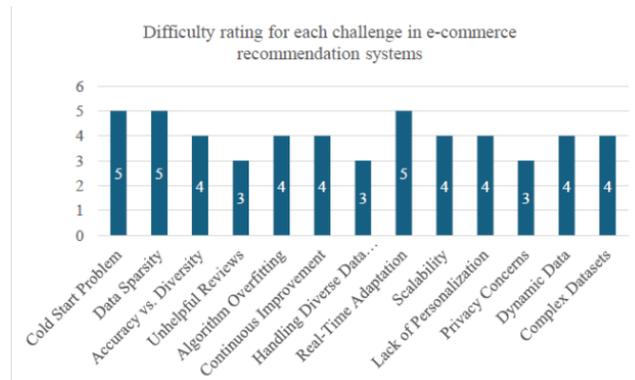

Figure 4: Visualization of primary challenges in e-commerce RSs and their corresponding difficulty rating for handling.

Hybrid systems such as the one proposed by [21] demonstrate superior handling of cold start problems and improved diversity in recommendations. However, their complexity and computational cost pose challenges for implementation. Similarly, [22] explored hybrid models using implicit ratings, achieving higher accuracy in niche markets and improved user satisfaction, but noted the dependency on large datasets and potential privacy concerns. [24]

emphasized the capability of hybrid collaborative filtering to handle sparse data effectively, leading to improved accuracy, albeit at the cost of extensive data preprocessing. Meanwhile, [28] introduced hybrid CNN-based models, which enhanced recommendation quality by filtering out unhelpful reviews but required significant computational resources.

The strengths of combining content-based and collaborative filtering methods were highlighted by [29], who achieved optimized accuracy but faced challenges related to potential overfitting and complexity of balancing algorithm weights. Industry examples, such as Netflix's hybrid recommendation system [34], underscore importance of continuously evolving algorithms for high business value, though such systems demand substantial resources for maintenance and updates. Innovative normalization techniques by [35] demonstrated improvements in handling diverse data sources, although these techniques remain complex to implement. Lastly, [37] presented dynamic hybrid approach that adapts to real-time customer behavior, resulting in higher user engagement, but this approach requires robust infrastructure and comes with high complexity.

### 4.4 RQ4

The field of recommendation systems in e-commerce has witnessed significant advancements with the integration of deep learning, reinforcement learning, and other cutting-edge technologies. These trends aim to address key challenges while optimizing user experience and business outcomes (Table 5).

## 5 Discussion and conclusions

This systematic literature review (SLR) analyzed 38 studies published between 2013 and 2025 to evaluate the effectiveness of machine learning methods in e-commerce recommendation systems (RSs). The analysis revealed that collaborative filtering, content-based filtering, and hybrid models are the most commonly used techniques in this domain. Hybrid models, which combine the strengths of both collaborative and content-based approaches, were found to outperform traditional methods in terms of handling diverse user needs and complex datasets. Additionally, integration of advanced ML techniques such as deep learning and reinforcement learning is becoming increasingly popular, offering improved scalability, personalization, and adaptability for e-commerce platforms.

Despite the promising results, the effectiveness of these methods varies depending on the specific e-commerce context and dataset used. Performance metrics such as accuracy, precision, and recall showed discrepancies across studies, highlighting the need for tailored approaches. Furthermore, scalability and the cold-start problem remain significant challenges for large-scale systems. SLR demonstrates that while significant progress has made, further advancements in hybrid and deep learning-based approaches are needed to address limitations in current recommendation systems.

The findings of this review have several practical implications for development of IT solutions in the e-commerce sector. For IT professionals and organizations, the insights into strengths and weaknesses of various recommendation system methods can help guide the selection of the most appropriate technology for specific business needs. Hybrid recommendation systems offer an opportunity to increase the precision and relevance of recommendations. Additionally, adoption of advanced techniques could provide substantial benefits in terms of scalability and personalized user experiences.

For organizations looking to implement or improve recommendation systems, it is essential to consider the infrastructure required for hybrid and deep learning-based methods, as these models can be complex and resource-intensive. Scalability remains a critical consideration, especially for large e-commerce platforms with large datasets. Addressing cold-start issues and integrating real-time feedback into the recommendation process can significantly enhance system effectiveness.

Based on the findings of this systematic literature review (SLR), several recommendations can be made for both practitioners and researchers in the field of e-commerce recommendation systems.

E-commerce businesses should consider adopting hybrid recommendation systems to improve the accuracy and relevance of product recommendations. Investing in deep learning and reinforcement learning can offer long-term benefits, particularly in terms of personalization and adaptability to evolving user preferences. However, careful attention should be given to the scalability and complexity of these systems, and appropriate infrastructure should be put in place. Future studies should focus on real-world, large-scale applications of hybrid and deep learning-based recommendation systems to better understand how these techniques perform in complex, dynamic e-commerce environments. Research on addressing scalability issues and cold-start problems should be prioritized, as these remain significant challenges for many e-commerce platforms. Additionally, exploring ethical concerns, such as data privacy and fairness, in the context of machine learning-based recommendation systems should be an area of future investigation.

In conclusion, while machine learning has greatly improved the effectiveness of e-commerce recommendation systems, there is still much to be explored and developed. Both practitioners and researchers must continue to innovate and address existing challenges to ensure that recommendation systems meet the growing demands of the e-commerce industry.

**Table 4: Comparison of hybrid RSs approach based on their performance attributes**

| Study | Approach | Accuracy | Diversity | Scalability | Real-Time Adaptability | Computational Cost | Complexity |
|---|---|---|---|---|---|---|---|
| [21] | Hybrid (Cold Start Aware) | High | High | Medium | Low | High | High |
| [22] | Hybrid (Implicit Ratings) | High | Medium | Low | Low | Medium | High |
| [24] | Hybrid Collaborative Filtering | High | Medium | Medium | Low | Medium | Medium |
| [28] | Hybrid CNN-Based | High | Low | Medium | Low | High | High |
| [29] | Hybrid (Content & Collaborative Filtering) | High | Medium | Medium | Low | Medium | Medium |
| [34] | Hybrid (Netflix Approach) | High | High | High | Medium | High | High |
| [35] | Hybrid (Normalization) | High | Medium | High | Low | Medium | High |
| [37] | Hybrid (Dynamic Analysis) | High | High | High | High | High | High |

**Table 5: Key advancements and trends shaping the future of RSs in e-commerce**

| Focus Area | Studies | Key Trends/Advancements |
|---|---|---|
| AI and Personalization | [3], [17], [14] | AI-driven personalization, Micro-behavioral analysis, Pre-trained language models |
| Scalability and Performance | [4] | Scalability improvements |
| Machine Learning & Deep Learning | [12], [9], [8] | Advanced machine learning algorithms, Deep learning clustering |
| Behavioral Analysis | [15], [17] | Behavioral sequence transformers, Micro-behavioral analysis |
| Hybrid Systems | [22], [24], [29] | Combination of multiple recommendation approaches |
| Reinforcement Learning | [19] | Reinforcement learning |
| Profit Maximization Models | [20] | Profit maximization models |